\documentclass[preprint,tightenlines,nofootinbib,superscriptaddress,floatfix]{revtex4}

\usepackage[bookmarks=false,hyperfootnotes=false]{hyperref}
\usepackage{graphicx}
\usepackage{amsmath}
\usepackage{xspace}

\newcommand{\eq}[1]{Eq.~\eqref{eq:#1}}
\newcommand{\fig}[1]{Fig.~\ref{fig:#1}}

\newcommand{\abs}[1]{\lvert#1\rvert}

\newcommand{\df}{\mathrm{d}}

\newcommand{\g}{\gamma}
\newcommand{\w}{\omega}

\newcommand{\hm}{\widehat{m}}
\newcommand{\hC}{\widehat{C}}
\newcommand{\hF}{\widehat{F}}
\newcommand{\hW}{\widehat{W}}

\newcommand{\GeV}{\,\mathrm{GeV}}

\newcommand{\nn}{\nonumber}

\newcommand{\lqcd}{\Lambda_\mathrm{QCD}}
\newcommand{\cut}{\mathrm{cut}}
\newcommand{\incl}{\mathrm{incl}}

\newcommand{\babar}{\mbox{\ensuremath{{\displaystyle B}\!{\scriptstyle A}{\displaystyle B}\!{\scriptstyle AR}}}\xspace}
\newcommand{\belle}{Belle\xspace}

\allowdisplaybreaks[2]

\begin{document}


\preprint{\vbox{\hbox{MIT--CTP 4206}\hbox{HU-EP-11/05}}}

\title{Status of SIMBA}

\author{Florian U. Bernlochner}
\affiliation{Humboldt University of Berlin, 12489 Berlin, Germany}

\author{Heiko Lacker}
\affiliation{Humboldt University of Berlin, 12489 Berlin, Germany}

\author{Zoltan Ligeti}
\affiliation{Ernest Orlando Lawrence Berkeley National Laboratory, University of California,\\ Berkeley, CA 94720, USA}

\author{\\Iain W. Stewart}
\affiliation{Center for Theoretical Physics, Massachusetts Institute of Technology,\\ Cambridge, MA 02139, USA}
\affiliation{Center for the Fundamental Laws of Nature, Harvard University,\\ Cambridge, MA 02138, USA}

\author{Frank J. Tackmann\footnote{Speaker}}
\affiliation{Center for Theoretical Physics, Massachusetts Institute of Technology,\\ Cambridge, MA 02139, USA}

\author{Kerstin Tackmann}
\affiliation{CERN, CH-1211 Geneva 23, Switzerland\vspace{5ex}}

\collaboration{The SIMBA Collaboration}

\begin{abstract}

The goal of the SIMBA collaboration is to provide a global fit to the available data in inclusive $B\to X_s\g$ and $B\to X_u\ell\nu$ decays. By performing a global fit one is able to simultaneously determine the relevant normalizations, i.e.\ the total $B\to X_s\g$ rate and the CKM-matrix element $\abs{V_{ub}}$, together with the required input parameters, most importantly the $b$-quark mass and the $b$-quark distribution function in the $B$-meson, called the shape function. This strategy is analogous to the determination of $\abs{V_{cb}}$ from global fits to inclusive $B\to X_c\ell\nu$ decays. In this talk, we present preliminary results for the shape function and $\abs{C_7^\incl V_{tb} V_{ts}^*}$, which parametrizes the total $B\to X_s\gamma$ rate, from a global fit to the available $B\to X_s\g$ measurements from \babar and \belle.

\vfill
\noindent\emph{Proceedings of CKM2010, the 6th International Workshop on the CKM Unitarity Triangle,\\
University of Warwick, UK, 6-10 September 2010}

\end{abstract}

\maketitle

\section{\boldmath A Global Fit Approach to $B\to X_s\g$ and $B\to X_u\ell\nu$}
\label{sec:intro}

The flavor-changing neutral current $B\to X_s\gamma$ process is very sensitive to contributions from new physics beyond the Standard Model (SM). To exploit this sensitivity and constrain new physics, an accurate extraction of the $B\to X_s\gamma$ rate from data is desirable. Currently, the experimentally measured partial branching fractions with a cut on the photon energy, $E_\g > E_\g^\cut$, are extrapolated to the partial branching fraction for a fixed cut $E_\g > 1.6\GeV$, which yields~\cite{TheHeavyFlavorAveragingGroup:2010qj} $\mathcal{B}(E_\g > 1.6) =  (3.55 \pm 0.24 \pm 0.09) \times 10^{-4}$. This value is then compared to the fixed next-to-next-to-leading order (NNLO) SM prediction from Refs.~\cite{Misiak:2006ab, Misiak:2006zs}, $\mathcal{B}(E_{\g} > 1.6) =  (3.15 \pm 0.23) \times 10^{-4}$. This procedure is used to compare experiment and theory because fixed-order perturbation theory can only be applied for low enough values of $E_\g^\cut$. However, since the extrapolation down to $E_\g^\cut = 1.6\GeV$ uses theory, it still requires a theoretical calculation of the decay rate with the actual value of $E_\g^\cut$ used in each measurement. In other words, irrespectively of how the extrapolation is performed, the comparison of theory and experiment always happens effectively at the measured $E_\g^\cut$.

The currently performed extrapolation and average~\cite{TheHeavyFlavorAveragingGroup:2010qj} assumes a model for the leading shape function, which introduces a systematic uncertainty from the model dependence. The extrapolation also has an additional theory uncertainty which is correlated to some extent with that of the fixed-order SM prediction at $E_\g^\cut = 1.6\GeV$. To minimize the effect of the extrapolation, from each experimental analysis typically the measurement with the smallest possible $E_\g^\cut$ is used, which has the largest experimental systematic uncertainty. Hence, a small subset of the experimental information dominates the result, and in particular the more precise measurements at higher values of $E_\g$ cannot be utilized.

Our strategy, which was proposed in Ref.~\cite{Ligeti:2008ac}, avoids these drawbacks. Performing a global fit allows one to minimize the uncertainties by making maximal use of all available data at any $E_\g$. At the same time it allows for a consistent treatment of correlated uncertainties, both experimental and theoretical, as well as from input parameters. We use a model-independent treatment of the shape function, such that its shape and uncertainty is determined by the shape and uncertainties in the measured $E_\g$ spectra. The overall $b\to s\g$ transition rate, which holds the sensitivity to new physics, is parametrized by the combination $\abs{C_7^\incl V_{tb} V_{ts}^*}$, defined below, and is determined by the normalization of the measured spectra. Its value obtained from the global fit can then be compared to its SM prediction in order to constrain possible contributions from new physics. In this way, measurements at all $E_\g$ contribute optimally to constrain the total $B\to X_s\g$ rate. Furthermore, the extracted shape function provides a necessary input for the determination of $\abs{V_{ub}}$ from inclusive $B\to X_u\ell\nu$ decays, e.g.\ via a combined global fit, which is left for future work.

\section{Theory}
\label{sec:theory}

\subsection{Treatment of the Shape Function}
\label{subsec:SF}

The shape function renormalized in $\overline{\mathrm{MS}}$, $S(\w, \mu)$, which enters the description of $B\to X_s\g$, can be factorized as~\cite{Ligeti:2008ac}
\begin{equation} \label{eq:S_construction}
S(\w, \mu) = \int\! \df k\, \hC_0(\w-k, \mu)\, \hF(k)
\,.\end{equation}
Here $\hC_0(\w, \mu)$ is the $\overline{\mathrm{MS}}$-renormalized $b$-quark matrix element of the shape-function operator calculated in perturbation theory, while $\hF(k)$ is the nonperturbative contribution to $S(\w, \mu)$. The hats on $\hF(k)$, $\hC_0(\w, \mu)$, and $\hm_b$ below indicate that they are defined in a short-distance scheme. Here we use the $1S$ scheme, $\hm_b \equiv m_b^{1S}$, see Ref.~\cite{Ligeti:2008ac} for more details.

The construction in \eq{S_construction} has several advantages. It ensures that $S(\w, \mu)$ has the correct perturbative tail at large $\w$ and also the correct $\mu$ dependence and RGE, which both come from $\hC_0(\w, \mu)$. For small $\w$, the shape of $S(\w, \mu)$ is determined by $\hF(k)$. Hence, $\hF(k)$ is the nonperturbative parameter that determines the shape of the $B \to X_s\g$ spectrum at large $E_\g$ and which we need to extract from the data. For simplicity, we will refer to $\hF(k)$ as the shape function in the following. In contrast to $S(\w, \mu)$, $\hF(k)$ falls off exponentially at large $k$, so the moments of $\hF(k)$ exist without a cutoff, and information about $m_b$ and matrix elements of local operators can be incorporated via constraints on the moments of $\hF(k)$. For example,
\begin{equation}
\int\!\df k\, \hF(k) = 1
\,,\qquad
\int\!\df k\, k \hF(k) = m_B - \hm_b
\,.\end{equation}

To fit $\hF(k)$ from data we follow Ref.~\cite{Ligeti:2008ac} and expand it in a complete orthonormal basis,
\begin{equation} \label{eq:basis}
\hF(k) = \frac{1}{\lambda} \biggl[ \sum_{n=0}^\infty \, c_n \, f_n\Bigl(\frac{k}{\lambda}\Bigr) \biggr]^2
\qquad \text{with} \qquad
\int \! \df k \, \hF(k)  = \sum_{n=0}^{\infty} c_n^2 = 1
\,.\end{equation}
The basis functions $f_n(x)$ are given in Ref.~\cite{Ligeti:2008ac} and $\lambda \simeq \lqcd$ is a dimension-one parameter of the basis. Since the functional basis is complete, \eq{basis} provides a model-independent description of $\hF(k)$, where its shape is parametrized by the basis coefficients $c_n$.

By fitting the coefficients $c_n$ from data, the experimental uncertainties and correlations in the measured spectra are captured in the uncertainties and correlations of the fitted $c_n$. In practice, the data only allow a fit to a finite number of coefficients, so the expansion must be truncated after $N+1$ terms. This introduces a residual model dependence from the chosen functional basis, in particular the value used for $\lambda$. The overall size of this truncation uncertainty scales as $1 - \sum_{n=0}^N c_n^2$. The optimal values for $\lambda$ and $N$ are determined from the data. The value for $\lambda$ is chosen such that the fitted series converges quickly, and the number $N$ of fit coefficients should be large enough such that the truncation uncertainty is small compared to the experimental uncertainties of the fit coefficients. In other words, we let the available data determine the precision to which the functional form of the shape function is known, by including as many basis coefficients in the fit as possible given the available data. Hence, our approach allows for an experimental determination of the shape function which is model independent and yields reliable, data-driven uncertainties.

\subsection{\boldmath Master Formula for $B\to X_s\gamma$}
\label{subsec:BtoXsg}

The $B \to X_s\g$ photon energy spectrum is given by
\begin{align} \label{eq:master}
\frac{ \df \Gamma}{\df E_\g}
&= \frac{G_F^2 \alpha_\mathrm{em}}{2\pi^4}\, E_\g^3\, \hm_b^2\,\lvert V_{tb} V_{ts}^* \rvert^2
\nn\\ &\quad\times
\biggl\{
\abs{C_7^\incl}^2 \biggl[\int\! \df k\, \hW_{77}(k) \hF(m_B - 2E_\g - k)
+ \sum_m  \hW_{77,m}\, \hF_m(m_B - 2E_\g) \biggr]
\nn\\ &\qquad
+  \int\! \df k \sum_{i,j\neq7}  \Bigl[2\mathrm{Re}(C_7^\incl) C_i\, \hW_{7i}(k)
+ C_i C_j\, \hW_{ij}(k) \Bigr] \hF(m_B - 2E_\g - k)
\biggr\}
\,.\end{align}
The expressions entering Eq.~\eqref{eq:master} will be given in Ref.~\cite{Ligeti2010}.
The function $\hW_{77}(k)$ contains the perturbative corrections to the $b\to s\gamma$ decay via the electromagnetic dipole operator, $O_7$, resummed to next-to-next-to-leading-logarithmic order~\cite{Becher:2006pu, Ligeti:2008ac}, and including the full NNLO corrections~\cite{Melnikov:2005bx, Blokland:2005uk}. At lowest order, $W_{77}(k) = \delta(k)$. The $\hF_m(k)$ are $1/m_b$ suppressed subleading shape functions. In a fit to $B \to X_s \g$ data only, they can be absorbed into $\hF(k)$ at lowest order in $\alpha_s$. The terms proportional to $C_{i\neq 7}$ are due to IR-finite bremsstrahlung corrections from operators other than $O_7$. They are included at next-to-leading order (NLO) for $i = 1,2,8$ using the SM values for $C_{1,2,8}$. They have almost no effect on the fit, because they are very small in the experimentally accessible region of the photon energy spectrum.

The coefficient $C_7^\incl$ multiplying the dominant $77$ contribution in \eq{master} is defined as
\begin{align} \label{eq:C7incl}
C_7^\incl
&= C_7^\mathrm{eff}(\mu_0) \frac{\overline m_b(\mu_0)}{\hm_b} + \sum_{i=1}^6 r_i(\mu_0)\, C_i(\mu_0)
+ r_8(\mu_0)\, C_8^\mathrm{eff}(\mu_0) \frac{\overline m_b(\mu_0)}{\hm_b}
+ \dotsb
\,.\end{align}
Here, $C_i^\mathrm{eff}(\mu_0)$ are the standard scheme-independent effective Wilson coefficients and $\overline m_b(\mu_0)$ is the $\overline{\mathrm{MS}}$ $b$-quark mass. The coefficients $r_{1-6,8}(\mu_0)$ contain all virtual contributions from the operators $O_{1-6,8}$ that generate the same effective $b\to s\g$ vertex as $O_7$. The ellipses denote included terms proportional to $\ln(\mu_0/\hm_b)$ that are required to cancel the $\mu_0$ dependence on the right-hand side and vanish at $\mu_0 = \hm_b$, such that $C_7^\mathrm{incl}$ is by definition $\mu_0$-independent to the order one is working at.

Since the terms in the last line in \eq{master} are small, we can consider $\lvert C_7^\mathrm{incl}\, V_{tb} V_{ts}^* \rvert$ as the parameter that parametrizes the normalization of the $B\to X_s\g$ rate. It is extracted simultaneously with $\hF(k)$ from our fit to the measured $E_\g$ spectra. The important contributions from $O_{1-6,8}$ are the virtual corrections contained in $C_7^\incl$, which have a sizable effect on the normalization of the $B\to X_s\g$ rate. By including them in $C_7^\incl$, they explicitly do not affect the shape of the spectrum, and so do not enter in our fit. They instead enter in the SM prediction for $C_7^\incl$, which can be computed independently. Below, we compare to the NLO SM value, $C_7^\incl = 0.354^{+0.011}_{-0.012}$~\cite{Ligeti2010}. For a more stringent test for new physics, evaluating $C_7^\incl$ in the SM at NNLO along the lines of Refs.~\cite{Misiak:2006ab, Misiak:2006zs} would be very valuable.

\section{\boldmath Fit to $B\to X_s\gamma$}
\label{sec:fit}

\vspace{-0.5ex}
\subsection{Fit Setup}
\vspace{-0.5ex}
\label{subsec:setup}

To fit to the experimentally measured photon energy spectra, we insert the expansion for $\hF(k)$ in \eq{basis} into \eq{master} and integrate over the appropriate range of $E_\g$ for each experimental bin and each combination of basis functions $f_m(x) f_n(x)$. The theory prediction for the $i$th bin, $B^i$, is then given by,
\begin{equation} \label{eq:fit}
B^i = \hm_b^2\, \abs{C_7^\incl V_{tb} V_{ts}^*}^2 \sum_{m,n=0}^N c_m c_n B_{mn}^i + \dotsb
\,,\end{equation}
where the ellipses denote the additional included terms arising from the last line in \eq{master}. The overall $\hm_b^2$ is expressed in terms of the moments of $\hF(k)$, so it is effectively a function of the $c_n$. We then perform a $\chi^2$ fit to all available bins with $c_{0,1,...,N}$ and $\abs{C_7^\incl V_{tb} V_{ts}^*}$ as the fit parameters. We enforce the constraint $c_0^2 + \dotsb + c_N^2 = 1$ to ensure that $\hF(k)$ is properly normalized to unity. (An estimate of the truncation uncertainty can then be obtained from the size of the last coefficient.)

As experimental inputs we use the \belle measurement from Ref.~\cite{:2009qg}, and the two \babar measurements from Refs.~\cite{Aubert:2007my,Aubert:2005cua}. The experimental statistical and systematic uncertainties and correlations are fully included in our fit. The \babar spectra are measured in the $B$ rest frame and are corrected for efficiencies. The experimental resolution in $E_\g$ for each spectrum is smaller than its respective bin size, so we can directly use both spectra in the fit. The \belle spectrum from Ref.~\cite{:2009qg} is measured in the $\Upsilon(4S)$ frame and affected by both efficiency and resolution. Correcting the spectrum to the $B$ rest frame depends on the shape function. We therefore boost our theory predictions to the $\Upsilon(4S)$ frame. Since the unfolded spectrum has very large bin-by-bin correlations, we apply the experimental detector response matrix to our theory predictions and fit to the measured spectrum. We have extensively tested our fitting procedure using pseudo-experiments.

The matrices $B_{mn}^i$ in \eq{fit} also have theoretical uncertainties, e.g.\ from higher-order perturbative corrections. The corresponding theory uncertainties in the fit results are roughly of the same size as the experimental ones. They are not yet included in the results below.

\subsection{Results}
\label{subsec:results}

\begin{figure}[t!]
\hfill\includegraphics[width=0.33\textwidth]{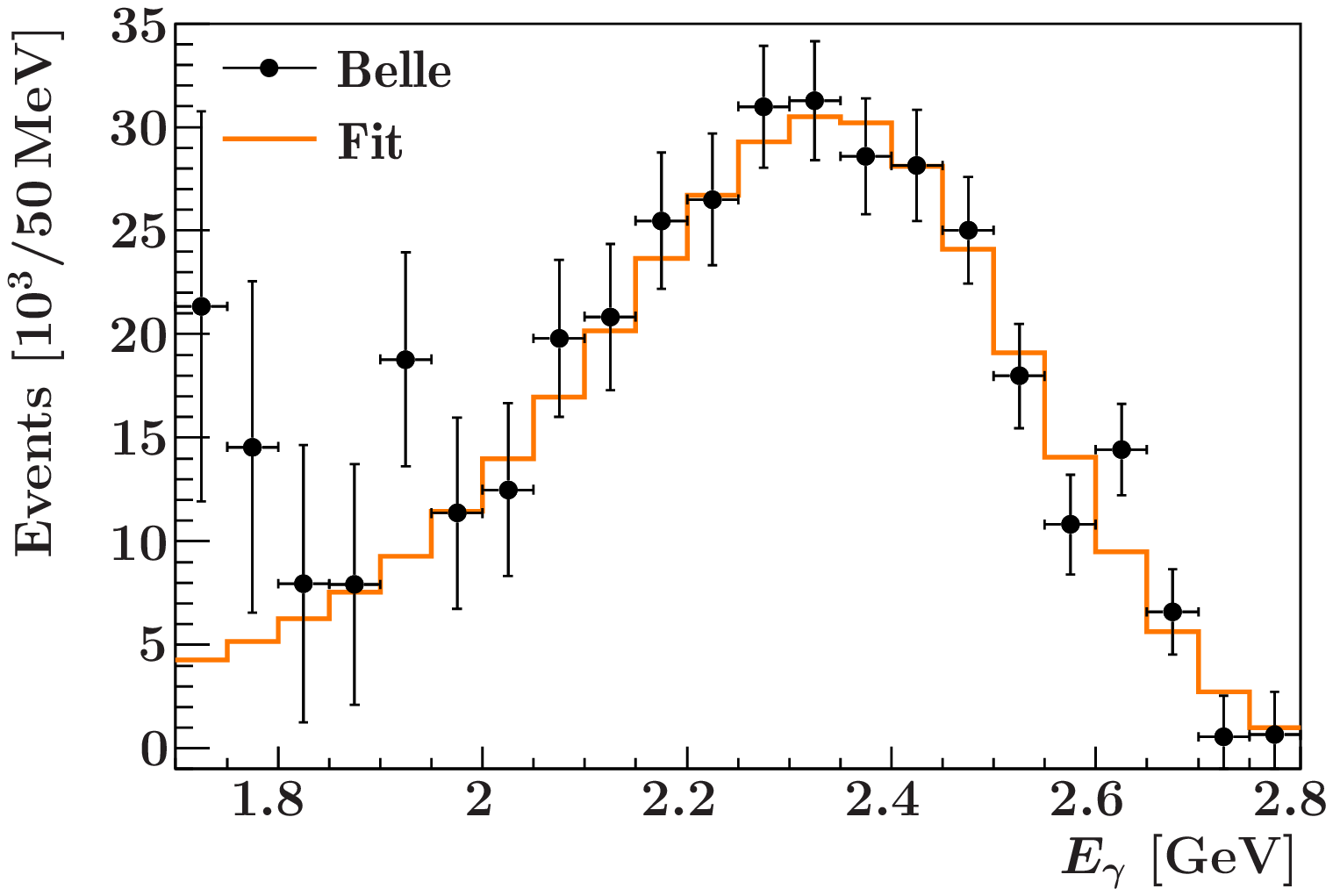}%
\hfill\includegraphics[width=0.33\textwidth]{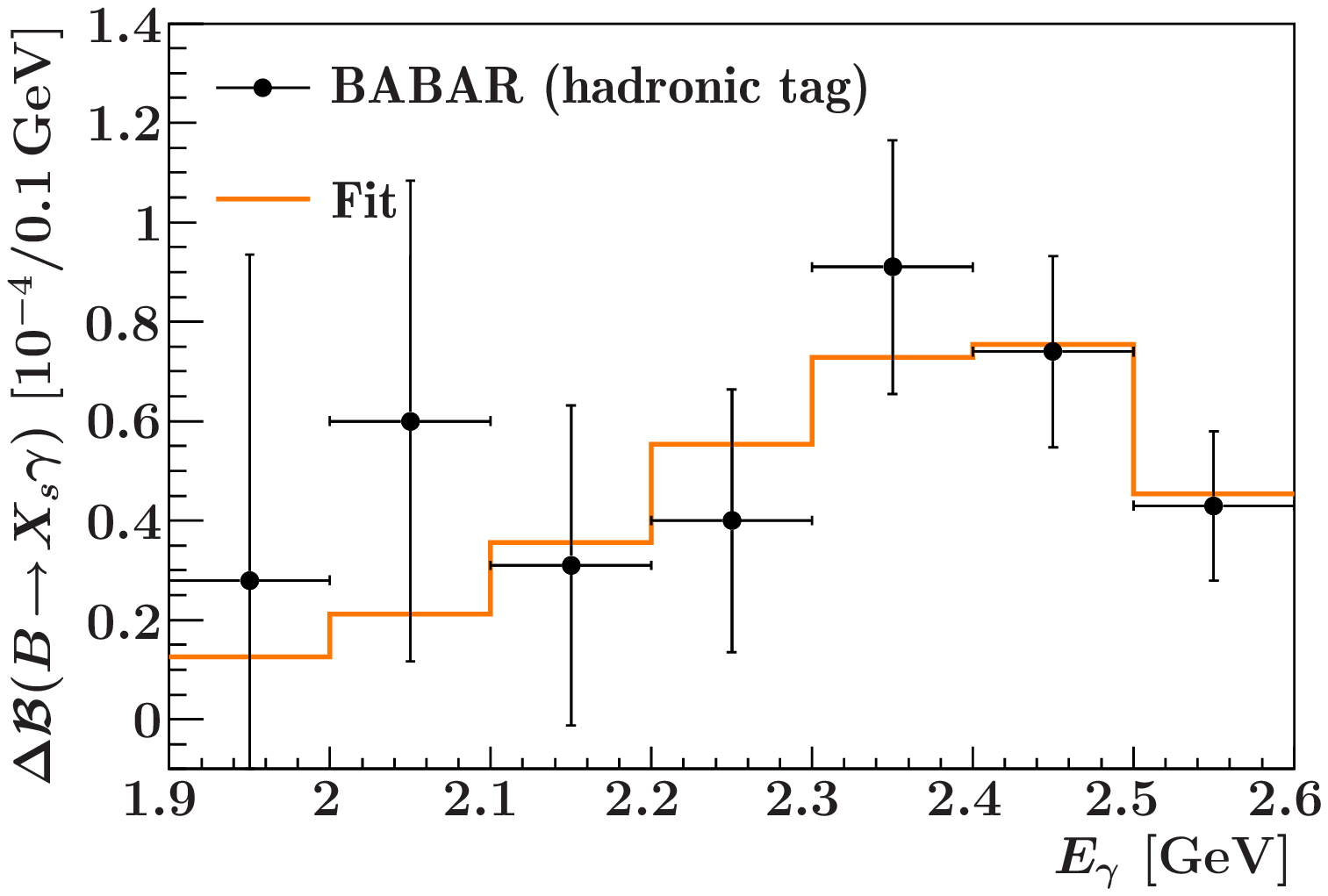}%
\hfill\includegraphics[width=0.33\textwidth]{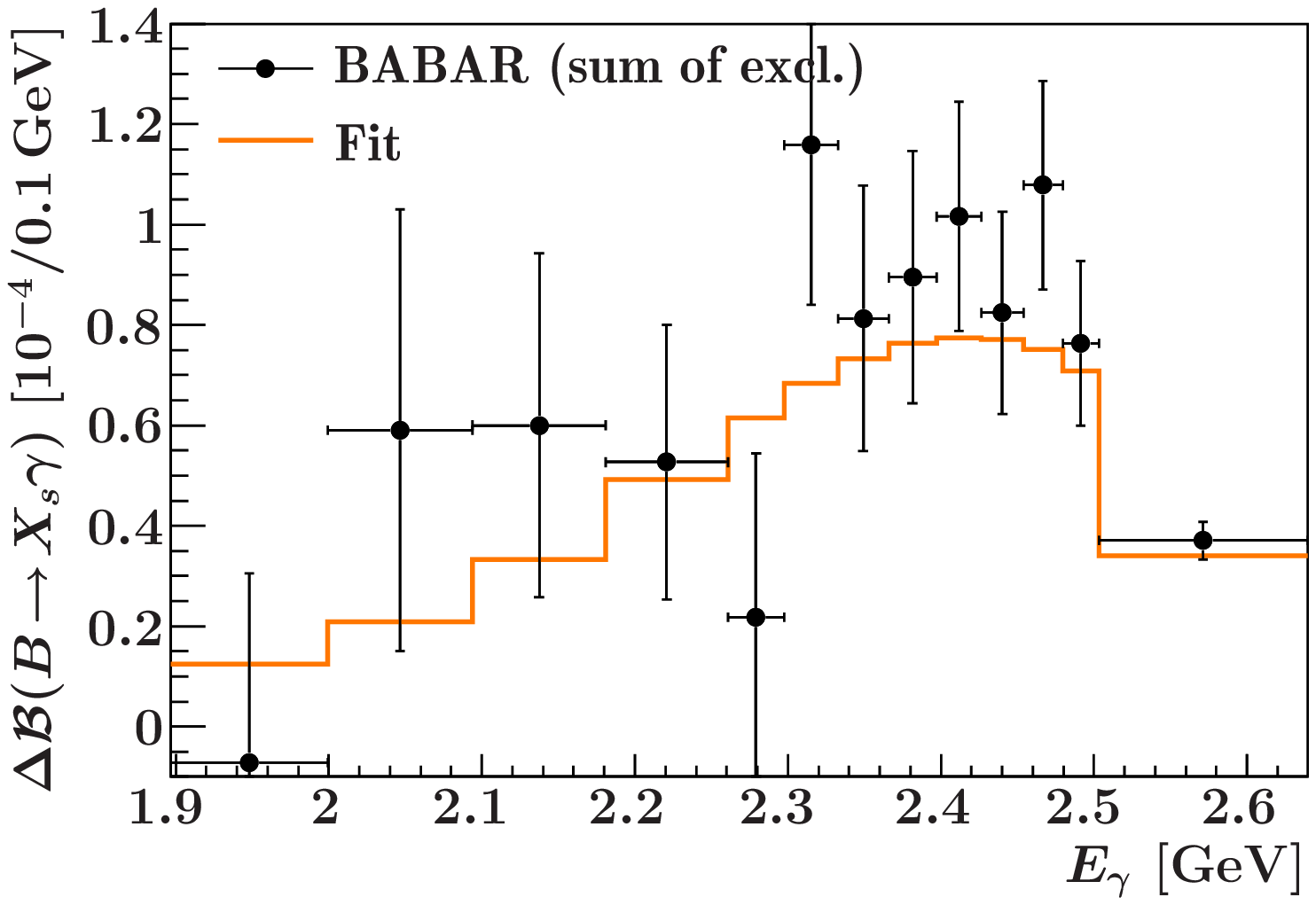}%
\hspace*{\fill}
\caption{Default fit to the \belle and \babar $B\to X_s\gamma$ photon energy spectra from Refs.~\cite{:2009qg,Aubert:2007my,Aubert:2005cua}. The points with error bars show the data, and the histograms show the fit result.}
\label{fig:fitresult}
\end{figure}

\begin{figure}[t!]
\parbox{0.5\textwidth}{\includegraphics[scale=0.5]{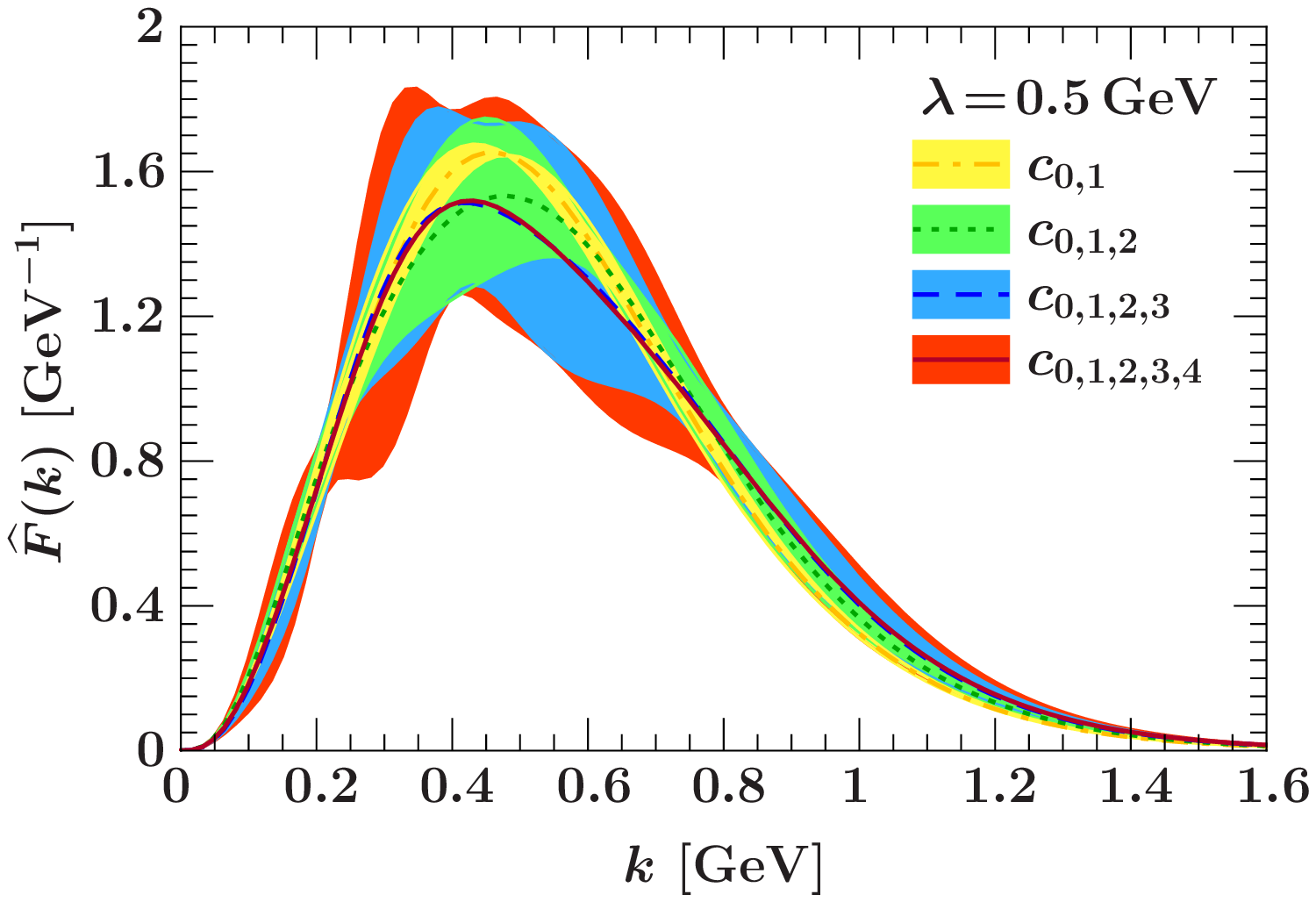}}%
\parbox{0.5\textwidth}{\includegraphics[scale=0.5]{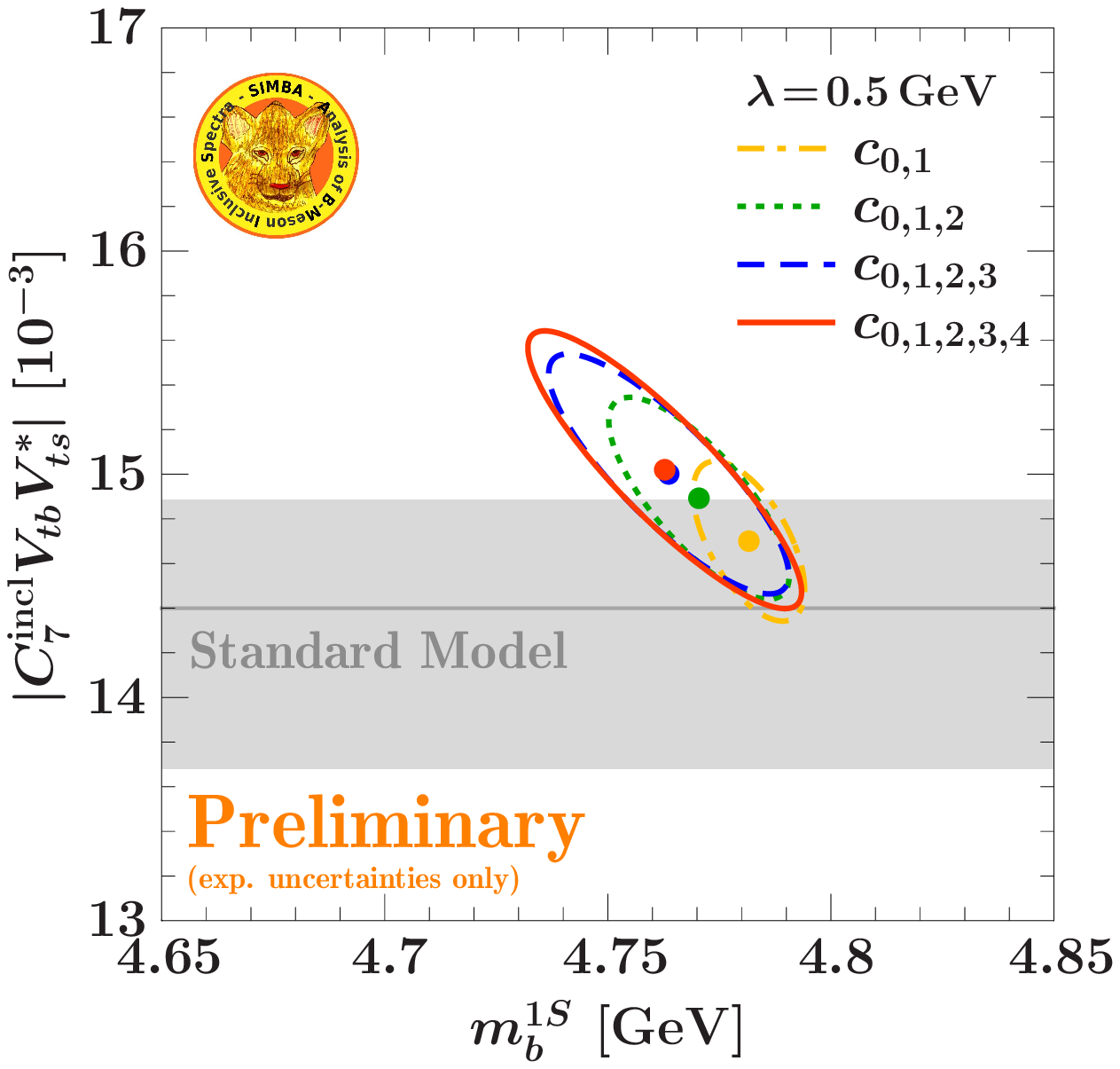}}%
\caption{Fit results for two ($c_{0,1}$), three ($c_{0,1,2}$), four ($c_{0,1,2,3}$), and five ($c_{0,1,2,3,4}$) basis coefficients and basis parameter $\lambda = 0.5\GeV$. Left: The extracted $\hF(k)$ (with absorbed $1/m_b$ corrections), where the colored envelopes are determined by the uncertainties and correlations of the fitted coefficients $c_n$. Right: The extracted values of $\lvert C_7^\mathrm{incl} \, V_{tb} V_{ts}^{*}\rvert$ and $m_b^{1S}$, where the colored ellipses show the respective $\Delta\chi^2 = 1$ contours, and the gray band shows the NLO SM value for $\lvert C_7^\mathrm{incl} \, V_{tb} V_{ts}^{*}\rvert$.}
\label{fig:SFC7mb}
\end{figure}

\begin{figure}[t!]
\parbox{0.5\textwidth}{\includegraphics[scale=0.5]{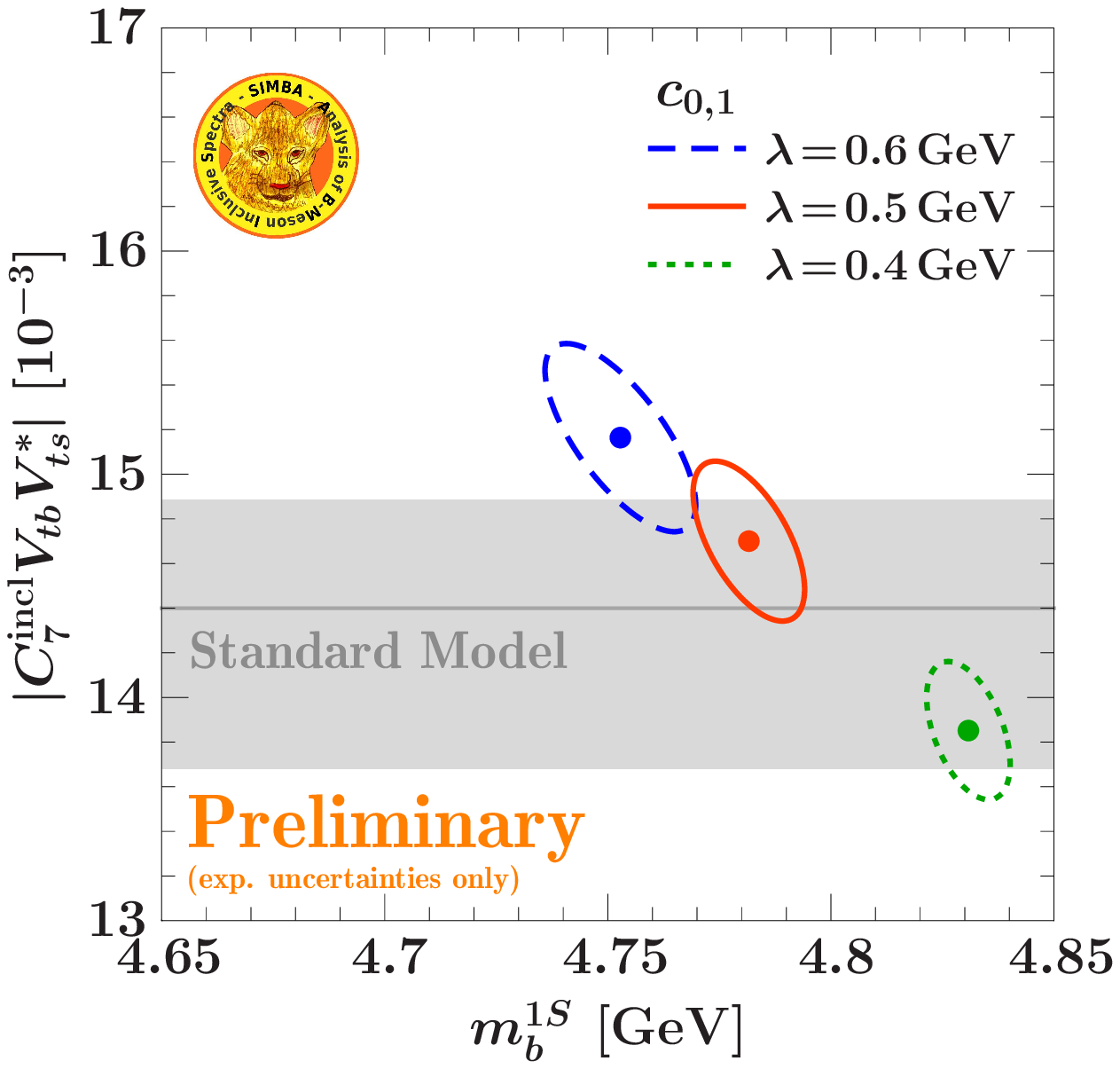}}%
\parbox{0.5\textwidth}{\includegraphics[scale=0.5]{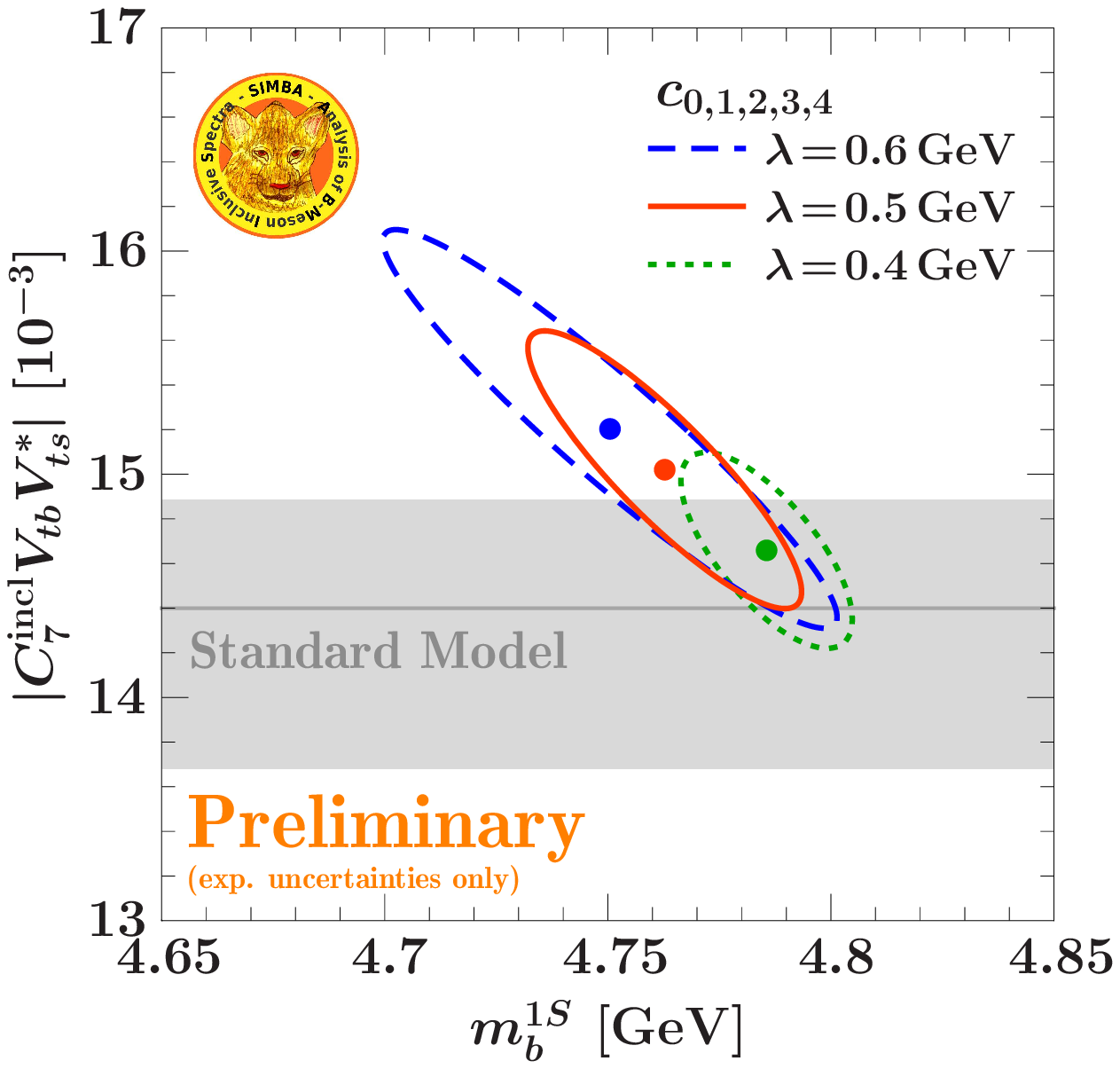}}%
\caption{Comparison of the fit results using different basis parameters using only two basis coefficients (left) and five basis coefficients (right).}
\label{fig:la040506}
\end{figure}

The results shown here are equivalent to those in Ref.~\cite{Bernlochner:2010tt}. For our default fit we use $\lambda = 0.5\GeV$ as basis parameter and four basis coefficients $c_{0,1,2,3}$. The fit has a $\chi^2 / \mathrm{dof} = 27.67/38$ and describes the measured spectra very well, as seen in \fig{fitresult}. The fit results for the shape function using each of $N+1 = 2,3,4,$ and $5$ basis coefficients are shown in the left panel of \fig{SFC7mb}. The corresponding results for $\abs{ C_7^\incl V_{tb} V_{ts}^{*}}$ and $m_b^{1S}$, where the latter is computed from the moments of the fitted $\hF(k)$, are shown in the right panel of \fig{SFC7mb}. Our default fit yields
\begin{equation}
\abs{C_7^\incl V_{tb} V_{ts}^{*}} = \bigl( 15.00 \pm 0.54_\mathrm{[exp]} \bigr)\times 10^{-3}
\,.\end{equation}
This result is still preliminary and does not yet include theory uncertainties. It agrees within one standard deviation with the NLO SM value, for which we use $\abs{V_{tb} V_{ts}^{*}} = 40.68^{+0.4}_{-0.5} \times 10^{-3}$.

The results in \fig{SFC7mb} verify the convergence of the basis expansion as the number of basis functions is increased. As one expects, the uncertainties returned by the fit increase with more coefficients due to the larger number of degrees of freedom. However, with too few coefficients we have to add the truncation uncertainty. A reliable value for the final uncertainty is provided by the fitted uncertainty when the central values have converged and the respective last coefficients, here $c_3$ or $c_4$, are compatible with zero. At this point, the truncation uncertainty can be neglected compared to the fit uncertainties. Equivalently, the increase in the fit uncertainties from including the last coefficient that is compatible with zero effectively takes into account the truncation uncertainty. Using a fixed model function and fitting one or two model parameters would thus underestimate the true uncertainties in the shape function. This is also seen in \fig{la040506}, which shows the results for different basis parameters $\lambda$. The left plot shows the results using only two basis coefficients in the fits. The three fits all have a good $\chi^2/\mathrm{dof} < 1$, but disagree with each other. This shows that there is an underestimated uncertainty due to the basis (i.e.\ shape) dependence when fitting too few basis coefficients. The right plot shows the corresponding results using five basis coefficients in the fit. In this case, the results agree very well within the fit uncertainties.

\section{Conclusions and Outlook}
\label{sec:conclusions}

We presented preliminary results from a global fit to $B\to X_s\g$ data, which determines the total $B\to X_s\g$ rate, parametrized by $\abs{C_7^\incl V_{tb} V_{ts}^{*}}$, and the $B$-meson shape function within a model-independent framework. The value of $\abs{C_7^\incl V_{tb} V_{ts}^{*}}$ extracted from data agrees with the SM prediction within uncertainties. From the moments of the extracted shape function we determine $m_b^{1S}$. In the future, information on $m_b$ from other independent determinations can be included by a constraint on the shape function. The shape function extracted from $B\to X_s\g$ is an essential input to the determination of $\abs{V_{ub}}$ from inclusive $B\to X_u\ell\nu$ decays.

A combined fit to $B \to X_s \g$ and $B \to X_u\ell\nu$ data within our framework is in progress. It will allow for a simultaneous determination of $\abs{C_7^\incl V_{tb} V_{ts}^{*}}$ and $\abs{V_{ub}}$ along with the shape function with reliable uncertainties. In addition to a few branching fractions with fixed cuts, it is important to have measurements of the $B\to X_u\ell\nu$ differential spectra (including correlations), e.g.\ the lepton energy or hadronic invariant mass spectra. As for $B\to X_s\g$, fitting the differential spectra allows making maximal use of the $B\to X_u\ell\nu$ data, by letting them constrain the nonperturbative inputs and further reduce the associated uncertainties.

\begin{acknowledgments}
We are grateful to Antonio Limosani from \belle for providing us with the detector response matrix of Ref.~\cite{:2009qg}. We thank Francesca Di Lodovico from \babar, who provided us with the correlations of Ref.~\cite{Aubert:2005cua}. This work was supported in part by
the Director, Office of Science, Offices of High Energy and Nuclear Physics of the U.S.\ Department of Energy under the Contracts DE-AC02-05CH11231, DE-FG02-94ER40818, and DE-SC003916.
\end{acknowledgments}

\vspace{-1ex}
\bibliographystyle{../physrev4}
\bibliography{../simba}

\providecommand{\href}[2]{#2}\begin{thebibliography}{10}

\bibitem{TheHeavyFlavorAveragingGroup:2010qj}
Heavy Flavor Averaging Group Collaboration, D.~Asner {\em et~al.},
\newblock \href{http://arXiv.org/abs/arXiv:1010.1589}{arXiv:1010.1589}.

\bibitem{Misiak:2006ab}
M.~Misiak and M.~Steinhauser,
\newblock Nucl. Phys. B {\bf 764}, 62 (2007),
  [\href{http://arXiv.org/abs/hep-ph/0609241}{hep-ph/0609241}].

\bibitem{Misiak:2006zs}
M.~Misiak {\em et~al.},
\newblock Phys. Rev. Lett. {\bf 98}, 022002 (2007),
  [\href{http://arXiv.org/abs/hep-ph/0609232}{hep-ph/0609232}].

\bibitem{Ligeti:2008ac}
Z.~Ligeti, I.~W. Stewart, and F.~J. Tackmann,
\newblock Phys. Rev. D {\bf 78}, 114014 (2008),
  [\href{http://arXiv.org/abs/arXiv:0807.1926}{arXiv:0807.1926}].

\bibitem{Ligeti2010}
Z.~Ligeti, I.~W. Stewart, and F.~J. Tackmann,
\newblock \emph{Manuscript in preparation}.

\bibitem{Becher:2006pu}
T.~Becher and M.~Neubert,
\newblock Phys. Rev. Lett. {\bf 98}, 022003 (2007),
  [\href{http://arXiv.org/abs/hep-ph/0610067}{hep-ph/0610067}].

\bibitem{Melnikov:2005bx}
K.~Melnikov and A.~Mitov,
\newblock Phys. Lett. {\bf B620}, 69 (2005),
  [\href{http://arXiv.org/abs/hep-ph/0505097}{hep-ph/0505097}].

\bibitem{Blokland:2005uk}
I.~R. Blokland, A.~Czarnecki, M.~Misiak, M.~Slusarczyk, and F.~Tkachov,
\newblock Phys. Rev. D {\bf 72}, 033014 (2005),
  [\href{http://arXiv.org/abs/hep-ph/0506055}{hep-ph/0506055}].

\bibitem{:2009qg}
Belle Collaboration, A.~Limosani {\em et~al.},
\newblock Phys. Rev. Lett. {\bf 103}, 241801 (2009),
  [\href{http://arXiv.org/abs/arXiv:0907.1384}{arXiv:0907.1384}].

\bibitem{Aubert:2007my}
BABAR Collaboration, B.~Aubert {\em et~al.},
\newblock Phys. Rev. D {\bf 77}, 051103 (2008),
  [\href{http://arXiv.org/abs/arXiv:0711.4889}{arXiv:0711.4889}].

\bibitem{Aubert:2005cua}
BABAR Collaboration, B.~Aubert {\em et~al.},
\newblock Phys. Rev. D {\bf 72}, 052004 (2005),
  [\href{http://arXiv.org/abs/hep-ex/0508004}{hep-ex/0508004}].

\bibitem{Bernlochner:2010tt}
SIMBA Collaboration, F.~U. Bernlochner {\em et~al.},
\newblock PoS {\bf ICHEP2010}, 229 (2010),
  [\href{http://arXiv.org/abs/arXiv:1011.5838}{arXiv:1011.5838}].

\end{thebibliography}

\end{document}